\documentclass[amsmath,amssymb,twocolumn,showpacs,reprint,aps,superscriptaddress]{revtex4-1}
\usepackage{graphicx}
\usepackage{multirow}
\usepackage{dcolumn}
\usepackage{bm}
\newcommand {\eqrf}[1]{(\ref{#1})}

\newcounter{No}
\renewenvironment{enumerate}[1][(\roman{No})]
{\begin{list}{#1}{\topsep=0pt\parsep=0pt plus 1pt\itemsep=\parsep\leftmargin=0pt\itemindent=\parindent\usecounter{No}}\addtolength{\itemindent}{\labelwidth}}{\end{list}}

\begin{document}
\title{Does neutron transfer with positive Q-values enhance sub-barrier fusion cross section?}

\author{V.~A. Rachkov}
\affiliation{Flerov Laboratory of Nuclear Reactions, JINR, Dubna, 141980 Russia}
\author{A.~V. Karpov}
\affiliation{Flerov Laboratory of Nuclear Reactions, JINR, Dubna, 141980 Russia}
\author{A.~S. Denikin}
\affiliation{International University "Dubna", Dubna, 141980 Russia}
\affiliation{Flerov Laboratory of Nuclear Reactions, JINR, Dubna, 141980 Russia}
\author{V.~I. Zagrebaev}
\affiliation{Flerov Laboratory of Nuclear Reactions, JINR, Dubna, 141980 Russia}

\begin{abstract}
\begin{description}
    \item[Background]{Significant enhancement of sub-barrier fusion cross sections owing to neutron rearrangement with positive $Q$-values were found for many combinations of colliding nuclei. However several experimental results on fusion reactions were reported recently in which such enhancement has not been observed in spite of a possibility for neutron rearrangement with positive $Q$-values.}
    \item[Purpose]{We aim to clarify much better the mechanism of neutron rearrangement in sub-barrier fusion reactions to find the other requirements (beside positive $Q$-value) which favour (or prevent) sub-barrier fusion enhancement.}
    \item[Method]{Channel coupling approach along with the semi-classical model for neutron transfer have been used for analysis of available experimental data.}
    \item[Results]{(1) Only 1n and 2n transfers with positive $Q$-values have a noticeable impact on sub-barrier fusion. Positive $Q$-value for neutron rearrangement is necessary but not sufficient requirement for additional sub-barrier fusion enhancement takes place. (2) ``Rigidity'' of colliding nuclei in respect of collective excitations is important that the sub-barrier fusion enhancement due to neutron rearrangement with positive $Q$-value be clearly visible. (3) Neutron binding energy in ``donor'' nucleus has a strong impact only in the case of fusion of light weakly bound nuclei.
}
\end{description}
\end{abstract}
\maketitle

\section{Motivation}

Fusion enhancement below the Coulomb barrier is one of the intensively studied phenomena, that became possible owing to progress in experimental techniques. It is by now well established that to describe the fusion cross sections one needs to include coupling of relative motion to other degrees of freedom such as rotation of deformed nuclei and their surface vibrations. The sub-barrier fusion enhancement induced by surface deformations or rotation of heavy deformed nuclei is well understood and properly described within the quantum coupled channel (QCC) approach \cite{Dasso1987187, Hagino1999143, PhysRevC.65.011601, Zagrebaev_Samarin, PhysRevC.73.034606, NRV} and within the empirical coupled channel (ECC) model \cite{PhysRevC.64.034606}.

\begin{figure}[ht]
      \includegraphics[width=6cm]{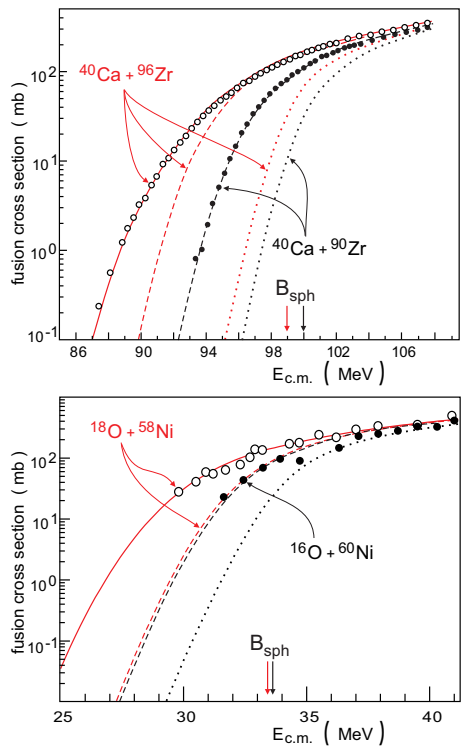}\\
      \caption{(Color online) (Upper panel) Fusion excitation functions for $^{40}$Ca+$^{96}$Zr (open circles) and $^{40}$Ca+$^{90}$Zr (filled circles) \cite{Timmers1998421}. The no-coupling limits are shown by the dotted curves. The dashed curves show the calculations with coupling to surface vibrations and without neutron transfer, whereas the solid line was obtained with accounting for neutron rearrangement in the entrance channel of the $^{40}$Ca+$^{96}$Zr reaction.
      (Bottom panel) Fusion excitation functions for $^{18}$O+$^{58}$Ni (open circles) and $^{16}$O+$^{60}$Ni (filled circles) \cite{Borges92}.
      The no-coupling limit is shown by dotted curve (practically indistinguishable for two reactions). Other notations are the same as in the upper panel.}
      \label{CaZr_ONi}
    \end{figure}

At the same time there are many experimental evidences testifying to additional enhancement of the sub-barrier fusion cross section due to neutron rearrangement with positive $Q$-values. This effect can be easier observed if one compares the sub-barrier fusion cross sections for two close projectile--target combinations for one of which neutron rearrangement with positive $Q$-values is possible whereas for another one all neutron transfers have negative $Q$-values. The combinations ($^{40}$Ca+$^{96}$Zr, $^{40}$Ca+$^{90}$Zr) \cite{Timmers1998421} and ($^{16}$O+$^{60}$Ni, $^{18}$O+$^{58}$Ni) \cite{Borges92} are good examples of such kind. Experimental and theoretical cross sections for these fusion reactions are shown in Fig.\ \ref{CaZr_ONi} taken from Ref.\cite{ZagrebaevPRC.67.061601}. Coupling of relative motion to the surface vibrations of target nuclei describes quite well the fusion cross sections for the $^{40}$Ca+$^{90}$Zr and $^{16}$O+$^{60}$Ni reactions, but it is insufficient to describe additional sub-barrier fusion enhancement for the $^{40}$Ca+$^{96}$Zr and $^{18}$O+$^{58}$Ni reactions caused by the intermediate neutron rearrangements with positive $Q$-values.

Another example of such kind is a comparison of the fusion cross sections for the $^{40}$Ca+$^{48}$Ca and $^{48}$Ca+$^{48}$Ca reactions \cite{PhysRevC.65.011601}. In this case quite unexpectedly sub-barrier fusion was found less probable for more neutron rich system. It is explained again by the possibility for neutron rearrangement with positive $Q$-value in the case of $^{40}$Ca+$^{48}$Ca but not in the case of $^{48}$Ca+$^{48}$Ca. The mechanism of sequential fusion was proposed in \cite{ZagrebaevPRC.67.061601} which described for the first time an additional sub-barrier fusion enhancement owing to neutron rearrangement with positive $Q$-value at approaching stage. The corresponding semiclassical model of this process was developed and successfully used for the predictions \cite{ZagrebaevPRC.67.061601, Adel2012119} and description of several fusion reactions \cite{PhysRevC.82.054609, Beck13, J.Phys:Conf_Ser420(2013)012124}.

Microscopic consideration of the problem performed within the time-dependent Schr\"{o}dinger equation \cite{PhysRevC.75.035809} and in the time-dependent Hartree-Fock (TDHF) approach \cite{JPhConfSer_420(2013)012118} clearly demonstrates that spreading of the valence neutron's wave function into the volume of the other nucleus takes place {\it before} the colliding nuclei overcome the Coulomb barrier. Thus, neutron rearrangement at approaching stage may really influence the sub-barrier fusion dynamics giving a gain in kinetic energy of colliding nuclei if it occurs with positive $Q$-value.

Last years the study of fusion reactions involving light weakly bound nuclei have been of increased interest \cite{springerlink:10.1007/BF01289520, PhysRevLett.81.4580, PhysRevLett.84.2342, PhysRevC.70.044601,Raabe2004, PhysRevLett.96.162701, Keeley2007579, PhysRevC.80.054609, JPhys.282.012014, PhysRevLett.107.092701,Wolski2011}. For these nuclei coupling to surface vibrations and rotation of heavy target is less important because of their smaller size. However just the rearrangement of nucleons at approaching stage may lead to significant sub-barrier fusion enhancement owing to large positive $Q$-value. For example, deep sub-barrier fusion cross section of $^{6}$He with $^{206}$Pb was predicted to be several orders of magnitude larger as compared with fusion of  $^{4}$He+$^{208}$Pb \cite{ZagrebaevPRC.67.061601} which was confirmed later in experiments \cite{PhysRevLett.96.162701,Wolski2011}. However there is the quantitative inconsistency between the results of both measurements as well as the calculated cross section \cite{ZagrebaevPRC.67.061601} at sub-barrier region. In spite of the fact that a large enhancement of sub-barrier fusion follows from all the experiments \cite{PhysRevLett.96.162701,Wolski2011}, further experimental and theoretical clarification of the problem is required. Moreover, deep sub-barrier fusion of light nuclei (including exotic ones) may have also an impact for astrophysical nucleosynthesis \cite{PhysRevC.75.035809}.

It turns out that the role of neutron transfer in sub-barrier fusion reactions is not completely clear. Recently several projectile--target combinations were reported (for example, $^{58,64}$Ni+$^{130}$Te \cite{PRL.107.202701}, $^{16,18}$O+$^{76,74}$Ge \cite{PRC.86.044621}, $^{60,64}$Ni+$^{100}$Mo \cite{Fusion_Ni_Mo}) for which the measured fusion cross sections do not demonstrate noticeable enhancement at sub-barrier energies in spite of positive $Q$-values for neutron rearrangement. It is important to note that the semiclassical model \cite{ZagrebaevPRC.67.061601}, which takes into account neutron rearrangement, describes perfectly all these new data and also does not predict any fusion enhancement for these combinations (see below). This means that positive $Q$-value for neutron rearrangement is necessary but not sufficient requirement for sub-barrier fusion enhancement takes place.

The main purpose of this paper is to find the other factors (conditions) of neutron rearrangement with positive $Q$-value which favour (or prevent) sub-barrier fusion enhancement. To find these conditions we clarified more dipper the mechanism of neutron rearrangement in fusion reactions.
Our findings can be formulated briefly as follows. (1) Only 1n and 2n transfer with positive $Q$-values play a noticeable role in sub-barrier fusion of heavy ions.
(2) ``Rigidity'' of colliding nuclei in respect of collective excitations is important that the sub-barrier fusion enhancement due to neutron rearrangement with positive $Q$-value can be observed. (3) Neutron binding energy in ``donor'' nucleus has a strong impact on the transfer probability for light weakly bound nuclei. Basing on the obtained results we made some predictions and proposed several combinations of colliding nuclei with large effect of neutron rearrangement for the experimental study.

\section{The model}

It is rather difficult to include the nucleon transfer channels to the rigorous quantum channel-coupled approach.
The problem appears when, following the standard channel-coupled method, the total wave function is decomposed over collective
(rotation and/or vibrational) states and simultaneously over neutron transfer states.
In such a decomposition overcomplete and non-orthogonal basis functions are used, that requites special complicated technique or some simplifying assumptions.
Moreover in medium mass and heavy nuclei single particle states are spread over numerous exited states of these nuclei
(with appropriate spectroscopic factors) which hardly can be included in any microscopic CC scheme.

At the same time, the neutron rearrangement was quite consistently incorporated into the ECC approach \cite{PhysRevC.64.034606} using semiclassical approximation for the transfer probability \cite{ZagrebaevPRC.67.061601}. This method is not fully microscopic, but it takes into account approximately the main effects of neutron rearrangement with positive $Q$-values.
The ECC model with neutron rearrangement has been already successfully used in several papers \cite{ZagrebaevPRC.67.061601, PhysRevC.82.054609, Adel2012119, Beck13, J.Phys:Conf_Ser420(2013)012124} to reproduce and predict cross sections for sub-barrier fusion reactions of stable nuclei.

In fusion reactions of light and medium mass nuclei it was found that compound nucleus (CN) is formed right away colliding nuclei come in contact.
However with increasing the masses of fusing nuclei (for example, it concerns reactions leading to formation of superheavy nuclei) the quasi-fission process starts to play more and more significant role. For such reactions the ECC and QCC models give the so-called ``capture cross section'', which is larger than the fusion cross section (formation of CN) by the quasi-fission cross section.
In this paper we consider fusion reactions of light and medium mass nuclei, for which the impact of the quasi-fission process is expected to be small.
Therefore, one may use the traditional notation ``fusion cross section'' for the cross section calculated within the empirical (or quantum) CC model.

Collision dynamics of two nuclei is regulated mostly by the nucleus-nucleus potential consisting of the Coulomb, nuclear and centrifugal terms.
The contact point, $R_{\rm cont} = R_1+R_2$, is located at the shorter distance than the position of the Coulomb barrier, $R_{\rm cont}< R_B$.
Thus, the fusion probability (or cross section) is determined by the probability to pass through the potential barrier.
The fusion cross section is usually decomposed over the partial waves as
    \begin{equation}\label{Total_fusion_CS}
        \sigma_{fus} \left(E\right)=\frac{\pi \hbar^2}{2 \mu E}\sum_{l = 0}^{l_{cr}} (2l+1)T_l\left(E\right)
    \end{equation}
where $E$ is the center-of-mass energy, $\mu$ is the reduced mass of the system, $l$ is the orbital angular momentum, $l_{cr}$ is the lowest angular momentum
at which a pocket in interaction potential disappears, and $T_l(E)$ is the barrier penetration probability.
Approximating the radial dependence of the barrier by a parabola one can use the simple Hill-Wheeler formula \cite{Hill-WheelerPhysRev.89.1102}
for the penetration probability $T_l$
    \begin{gather}\label{Tl_HW}
        T_l^{HW}\left(B;E\right) = \notag \\ \left[1+\exp{\left(\frac{2\pi}{\hbar\omega_B\left(l\right)}\left[B + \frac{\hbar^2\, l(l+1)}{2\mu R_B^2(l)}-E\right]\right)}\right]^{-1},
    \end{gather}
where $B$ and $R_B(l)$ are the height and position of the potential barrier, respectively, $\hbar\omega_B\left(l\right)$ is the width of the parabolic barrier.

Generally the nucleus-nucleus interaction potential depends not only on the relative distance between the colliding nuclei, but also on their deformations ($\vec{\beta_i}$, $i=1,2$) and mutual orientations, $\theta_i$. Thus, the interaction potential is characterized by the multidimensional Coulomb barrier, height of which is a function of orientations and deformations: $B(\vec{\beta_1},\vec{\beta_2};\theta_1,\theta_2)$.
Within the ECC model \cite{PhysRevC.64.034606} the coupling of the radial motion to the collective degrees of freedom is taken into account by averaging the transmission coefficient over the dynamic surface deformations and/or orientations of deformed colliding nuclei.

Collision dynamics and, consequently, the fusion cross section for spherical nuclei depend mainly on the coupling to their surface vibration degrees of
freedom. Therefore, the partial penetration probability should be averaged over the deformation-dependent barrier height
    \begin{equation}\label{Tl_sph_nucl}
        T_l\left(E\right) = \int f(B) \;{T}_l^{HW}\left(B; E\right)dB,
    \end{equation}
where dynamic deformations are assumed to take place along inter-nuclear axis, and $f(B)$ is the empirical dynamic barrier distribution function
\cite{PhysRevC.64.034606} normalized to unity: $\int f(B)dB=1$. It is not the same as the (conventional) ``experimental'' barrier distribution function
$D(B) = {d^2}\left(E\cdot\sigma_{fus}\right)/{dE^2}$ \cite{Rowley199125}. For one dimensional (no-coupling) barrier model $f(B)=\delta(B-B_0)$
whereas $D(B)$ in this case is still a smooth function with one peak at $B=B_0$ and with a width of about $0.56\hbar\omega_B$ (see, for example, \cite{Zagrebaev_Samarin}).
For the realistic multi-dimensional barrier (simulating channel coupling) we use the Gaussian approximation for $f(B)$.
    \begin{equation}\label{FB}
        f(B) = N_B\cdot \exp\left(-\left[\frac{B-B_0}{\Delta B}\right]^2\right),
    \end{equation}
where $B_0 = (B_1+B_2)/2$. Here $B_1$ is the height of the barrier at zero dynamic deformation of colliding nuclei, $B_2$ is the height of the saddle point calculated with realistic vibrational properties of nuclei, i.e., with the surface stiffness parameters obtained from the experimental values of the excited vibrational states, $\Delta B = (B_1-B_2)/2$, and $N_B$ is the normalization coefficient.

In the case of collisions of statically deformed nuclei ($\beta_i^{g.s.} \neq 0$) the Coulomb barrier height and its position depend on mutual orientation
of nuclei, then the averaging over the orientations of both nuclei is required
    \begin{eqnarray}\label{Tl_def_nucl}
        T_l(E) = \frac{1}{4}\int\limits_0^{\pi} \int\limits_0^{\pi}
    {T}_l^{HW}\left(B(\vec{\beta}_1^{g.s.}, \vec{\beta}_2^{g.s.}; \theta_1, \theta_2 );
E\right)\times\notag \\ \sin \theta_1 \sin \theta_2\, d\theta_1\, d\theta_2.
    \end{eqnarray}
Note, that Eq. (\ref{Tl_def_nucl}) assumes uniform distribution over the initial orientations (the corresponding dynamic barrier distribution function is unity in the region of integration).

The neutron rearrangement channels can be easily included \cite{ZagrebaevPRC.67.061601} in this ECC model of fusion reactions.
The total penetration probability (which takes into account the rearrangement of neutrons) can be estimated again by formulas \eqrf{Tl_sph_nucl}
or \eqrf{Tl_def_nucl} in which ${T}_l^{HW}$ is replaced by the following expression
    \begin{eqnarray}
        \widetilde{T}_l^{HW}\left(B;E\right) = \frac{1}{N_{tr}}\sum_{x=0}^4 \int\limits_{-E}^{Q_{xn}} \alpha_k\left(E,l,Q\right)\times \notag\\ T_l^{HW}(B; E+Q)\;dQ, \label{Tl_HW_MOD}
    \end{eqnarray}
where $Q_{xn}$ is the $Q$-value of the ground-to-ground transfer of $x$ neutrons.
The probability of the transfer of $x$ neutrons with a given $Q$-value (less than $Q_{xn}$) is calculated as follows
    \begin{equation}
        \alpha_k\left(E,l,Q\right) = N_k\exp\left(-CQ^2\right)\exp\left(-2\kappa\left[D\left(E,l\right)-D_0 \right]\right),
    \label{ak}
    \end{equation}
where $\kappa  = \kappa (\varepsilon_1) + \kappa (\varepsilon_2)+...+\kappa(\varepsilon _k)$ for sequential transfer of $k$ neutrons, $\kappa(\varepsilon_i) = \sqrt{2\mu_n\varepsilon_i/{\hbar^2}}$ and $\varepsilon _i$ is the binding energy of the $i$-th transferred neutron, $D\left(E,l\right)$ is the distance of the closest approach along the Coulomb trajectory with angular momentum $l$, ${D_0} = R_1^{\left(n\right)} + R_2^{\left(n\right)} + d_0$, $R_i^{\left(n\right)}=r_0^{\left(n\right)}A^{1/3}$ are the orbit radii of the valence (transferred) neutrons of colliding nuclei ($r_0^{\left(n\right)}$ and $d_0$ are adjustable parameters), $N_{tr}$ is the normalization constant, $\alpha_0 = \delta (Q)$, $C = R_B\mu_{12}/4\kappa\hbar^2B$ and $\mu_{12}$ is the reduced mass of two nuclei in the entrance channel.

The values of $r_0^{\left(n\right)} = 1.25$~fm and $d_0 = 2.5$~fm were fixed to reproduce the experimental data for the fusion reactions such as ${}^{40,48}\text{Ca} + {}^{48}\text{Ca}$ \cite{PhysRevC.65.011601}, ${}^{40}\text{Ca} + {}^{90,96}\text{Zr}$ \cite{Timmers1998421}, ${}^{16,18}\text{O} + {}^{60,58}\text{Ni}$ \cite{PhysRevC.46.2360}, etc. Note that the values $r_0^{\left(n\right)} = 1.4$ and $d_0 = 0$ are usually extracted \cite{PhysRevC.84.034603,L_Corradi_G_Pollarolo} from the analysis of the data on transfer reactions. This leads to smaller values of $D_0$ (and smaller values of $\alpha_k$) for the transfer reactions as compared to those required for the fusion reactions. This difference can be understood because for the fusion reactions the effect from the neutron rearrangement depends on how strongly the wave function of valent neutron is spread over the two-center molecular states at the moment of closest approach, whereas final (measured) neutron transfer probability is given by the situation after the re-separation of the colliding nuclei at infinite distance between them.

As can be seen from (\ref{Tl_HW_MOD}) enhancement of the fusion probability may appear at sub-barrier energies if rearrangement of neutrons leads to a gain in energy (positive Q-values).
In the reactions with negative Q-values the neutron rearrangement in the entrance channel does not influence the total fusion cross section because the
penetration probability $T_l^{HW}(B; E+Q)$ becomes smaller for negative $Q$. In this case $\alpha_0$ is the only non-vanishing term in sum (\ref{Tl_HW_MOD}). Note, that the probability of the neutron rearrangement dependents not only on the Q-value but also on the binding energy of the
transferred neutron, see Eq.~(\ref{ak}). The coefficients $\alpha_k$ decrease fast with increase of the binding energies in the ``donor" nucleus.
However this effect usually ignored while discussing the influence of the neutron rearrangement on the fusion process.

In our calculations up to four neutron transfer channels are taken into account. However, due to fast decrease of $\alpha_k$ with increasing the number of transferred neutrons $k$, only 1n and 2n transfer channels with positive $Q$-values were found to play a significant role (see discussion below). The experiments indicate (see, for example, \cite{Z.Phys.A326(1987)463,L_Corradi_G_Pollarolo}) that simultaneous transfer of two neutrons might be enhanced by factor $N_{2n}\sim3$ as compared to independent (subsequent) transfer of these neutrons.

\section{Neutron transfers with positive Q-values do not always enhance sub-barrier fusion probability }

All calculations presented below have been performed (and can be easily repeated) with the NRV Fusion code allocated at the web site with free access \cite{NRV}. The Woods-Saxon potential with the parameters listed in Table \ref{tab_potential} is used as the nuclear part of the nucleus-nucleus interaction. The choice of the potential parameters is rather uncertain. Moreover, often the same behavior of the fusion cross sections can be obtained with few sets of the parameters. Therefore, to make the predictions for a chosen reaction reliable, first we fit the potential to reproduce available experimental data for the nearest projectile-target combination and then use the same parameters for the studied reaction.

    \begin{table}[t]
        \begin{center}
            \caption{Parameters of Woods-Saxon potentials, heights, positions, and curvatures of the corresponding fusion barriers}
            \label{tab_potential}
            \begin{tabular}{p{2cm}cccccc}
              \hline
              \hline
              \tabularnewline[-6pt]
              System & {$V_0$} & $r_0$ & $a$ & $B_{sph}$ & $R_B$ & $\hbar\omega_B$\\
                     &  MeV    & fm    & fm    & MeV   & fm   & MeV\\
              \tabularnewline[-6pt]
              \hline
              \rule{0pt}{15pt}
              $^{40}\text{Ca} + ^{90}\text{Zr}$ & proximity &  &  & 100.0 & 10.7 & 3.93\\
              \rule{0pt}{10pt}
              \tabularnewline[-8pt]
              \rule{0pt}{0pt}
              $^{40}\text{Ca} + ^{96}\text{Zr}$ & proximity &  &  & 99.0 & 10.8 & 3.90\\
              \rule{0pt}{10pt}
              \tabularnewline[-8pt]
              \rule{0pt}{0pt}
              ${}^{16}\text{O} + {}^{60}\text{Ni}$ & -54 & 1.11 & 0.55 & 34.1 & 8.8 & 4.20\\
              \rule{0pt}{10pt}
              \tabularnewline[-8pt]
              \rule{0pt}{0pt}
              ${}^{18}\text{O} + {}^{58}\text{Ni}$ & -54 & 1.12 & 0.55 & 33.5 & 9.0 & 3.96\\
              \rule{0pt}{10pt}
              \tabularnewline[-8pt]
              \rule{0pt}{0pt}
              $^{16}\text{O} + ^{154}\text{Sm}$ & -105 & 1.12 & 0.75 & 59.5 & 11.2 & 3.99\\
              \rule{0pt}{10pt}
              \tabularnewline[-8pt]
              \rule{0pt}{0pt}
              ${}^{58}\text{Ni} + {}^{130}\text{Te}$ & $-$108 & 1.12 & 0.76 & 169.0 & 11.5 & 3.64 \\
              \rule{0pt}{10pt}
              \tabularnewline[-8pt]
              \rule{0pt}{0pt}
              ${}^{64}\text{Ni} + {}^{130}\text{Te}$ & $-$108 & 1.1 & 0.76 & 169.6 & 11.4 & 3.52 \\
              \rule{0pt}{10pt}
              \tabularnewline[-8pt]
              \rule{0pt}{0pt}
              ${}^{16}\text{O} + {}^{76}\text{Ge}$ & $-$56.5 & 1.17 & 0.64 & 34.7 & 9.9 & 3.79 \\
              \rule{0pt}{10pt}
              \tabularnewline[-8pt]
              \rule{0pt}{0pt}
              ${}^{18}\text{O} + {}^{74}\text{Ge}$ & $-$56.5 & 1.14 & 0.62 & 35.6 & 9.7 & 3.78 \\
              \rule{0pt}{10pt}
              \tabularnewline[-8pt]
              \rule{0pt}{0pt}
              ${}^{60}\text{Ni} + {}^{100}\text{Mo}$ & $-$100 & 1.12 & 0.68 & 142.9 & 11.0 & 3.88\\
              \rule{0pt}{10pt}
              \tabularnewline[-8pt]
              \rule{0pt}{0pt}
              ${}^{64}\text{Ni} + {}^{100}\text{Mo}$ & $-$100 & 1.12 & 0.68 & 141.6 & 11.2 & 3.78 \\
              \rule{0pt}{10pt}
              \tabularnewline[-8pt]
              \rule{0pt}{0pt}
              ${}^4\text{He} + {}^{64}\text{Zn}$ & $-$110 & 0.9 & 0.65 & 9.7 & 8.2 & 4.26 \\
              \rule{0pt}{10pt}
              ${}^6\text{He} + {}^{64}\text{Zn}$ & $-$110 & 0.9 & 0.65 & 9.4 & 8.4 & 3.43 \\
              \rule{0pt}{10pt}
              \tabularnewline[-8pt]
             \hline
              \hline
            \end{tabular}
        \end{center}
    \end{table}

Coupling to target and projectile collective states were taken into account for each studied system. The parameters of the vibrational excitations for the QCC calculations are taken from the NRV experimental databases. For the parameters of dynamic deformations required for the ECC calculations we use experimental data for the energies of the lowest vibrational states and the liquid-drop values \cite{bohr1998:v2} of the stiffness parameters. To treat the excitation of rotational states for deformed nuclei we use the corresponding experimental data for the energy of the first rotational $2^{+}$ state and the ground state deformation parameters according to Ref. \cite{Moller1995}.

In each case a few close projectile-target combinations are considered, one of them does not reveal the influence of the neutron rearrangement and another one having the influence. As a rule the chosen combinations lead to the same compound nuclei in order to have the same decay properties that would simplify experiment itself and the analysis of the corresponding experimental data.
This is the reasonable way in the current studies when the collective properties of the fusing nuclei (first of all those of the target) are close. However if  collective properties of the targets are different this may lead to a large difference between sub-barrier fusion cross sections not owing to the neutron rearrangement but due to the different properties of collective states.

    \begin{figure}[ht]
      \includegraphics[width=7cm]{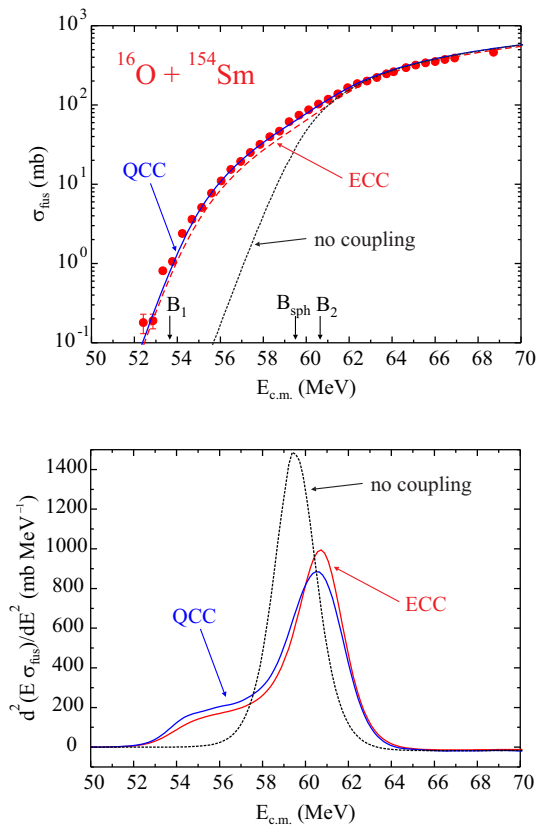}
      \caption{(Color online) Fusion cross section for $^{16}$O+$^{154}$Sm. The solid and dashed curves correspond to ECC and QCC calculations, respectively. Experimental data are from \cite{Leigh95}. Dotted curve shows the result for one-dimensional barrier penetrability. Arrows indicate the barriers for two limit orientations of deformed $^{154}$Sm nucleus ($B_1=53.6$ MeV and $B_2=60.7$ MeV) and for its spherical shape with the same volume, $B_{sph}$. At the bottom panel the corresponding ``experimental'' barrier distribution functions, $D(B)$, are shown.}
      \label{O_Sm}
    \end{figure}

First, we show that the ECC and QCC models give quite similar results for the systems where only the coupling to collective states play a role and not the neutron transfer.
Figure\ \ref{O_Sm} shows fusion cross sections in the reaction $^{16}$O+$^{154}$Sm \cite{Leigh95}. Coupling to rotational states of $^{154}$Sm ($E_{2+}=82$ keV, $\beta_2$ = 0.3, $\beta_4$ = 0.11) were included in the QCC and ECC calculations. The projectile is treated as a structureless nucleus. All the $Q$-values for the neutron rearrangement are negative in this reaction, and the neutron transfer channels do not influence the fusion cross sections. As can be seen the QCC and ECC approaches give very similar results. The same takes place for all other combinations of fusing nuclei. We use the ECC model below just because it allows one to include the coupling to neutron rearrangement channels by Eqs.(\ref{Tl_HW_MOD}) and (\ref{ak}).
    \begin{figure}[ht]
      \includegraphics[width=7cm]{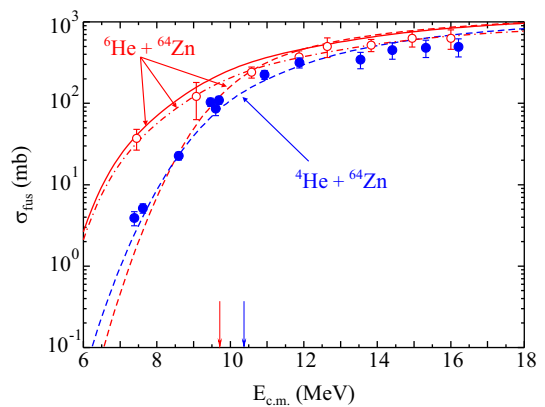}
      \caption{(Color online) Fusion cross section for $^{4,6}$He+$^{64}$Zn. The solid and dashed curves correspond to ECC calculations with and without accounting for the neutron transfer, respectively. Dash-dotted curve shows the $^{6}$He+$^{64}$Zn fusion cross section multiplied by the factor 0.8. Experimental data are from \cite{JPhys.282.012014}.}
      \label{He4_6_Zn}
    \end{figure}

It is known (see the discussion above) that the sub-barrier fusion of weakly bound nuclei is a ``classical'' example of the reaction revealing a strong enhancement due to neutron rearrangement. Figure \ref{He4_6_Zn} shows the fusion cross sections for two systems $^{4}$He+$^{64}$Zn (no neutron transfer, since all $Q_{xn}$ values are negative) and $^{6}$He+$^{64}$Zn having large $Q_{xn}$ values ($Q_{1n}=6.11$ MeV, $Q_{2n}=18.06$ MeV, and $Q_{3n}=4.54$ MeV). One may notice a good agreement with recent experimental data \cite{JPhys.282.012014} for the both systems. The ECC calculations without neutron rearrangement (dashed curves) are very close for two reactions. The enhancement owing to neutron rearrangement reaches of about one order of magnitude already at 1 MeV below the Coulomb barrier and it is even stronger at deeper sub-barrier energies.

In literature there are discussions of two counteractive factors influencing the fusion of weakly bound nuclei (such as the $^{6}$He one). First factor is owing to the neutron rearrangement with positive $Q$-values studied here. Second one is caused by the break-up of weakly bound projectile which is not included neither in the QCC nor in ECC calculations (the problem is studied in many paper, see, e.g., Refs.~\cite{Dasgupta2004,Mukherjee2006,PhysRevC.83.064606}). Coupling to the break-up and nucleon transfer channels reveals itself in nucleus-nucleus potential as the polarization terms having different signs. The neutron rearrangement with positive $Q$-values leads to the enhancement of sub-barrier fusion, i.e. it provides attractive polarization potential. The break-up processes result in repulsive additive to the potential and, therefore, suppress the fusion cross section (see, for example, Refs.~\cite{Yabana92,Matsumoto04}). In order to roughly estimate these two factors let us consider the fusion cross section at above barrier energies. It will be shown in Section \ref{Sect4} that the influence of the neutron rearrangement processes become weak with increasing the energies. Nevertheless the calculated cross section in Fig.~\ref{He4_6_Zn} overestimates experimental data. In the case of fusion reactions with the participation of light weakly bound nuclei like the $^{6}$He one such the damping of the cross section is usually attributed to the influence of the break-up channels. It plays noticeable role at near and above barrier energies and results in about 20 -- 30\% reduction of the magnitude of the fusion cross section \cite{Dasgupta2004,PhysRep.424.1}. Thus the calculated cross section can be easily fitted to the data being multiplied by empirical coefficient $\sim$ 0.7 -- 0.8 (dash-dotted curve in Fig.~\ref{He4_6_Zn}). At the deep sub-barrier energies which are in focus of our study the role of the break-up channels is much less while the influence of the transfer and inelastic channels dominate growing the cross section by the orders of magnitude. Therefore we do not use any additional coefficients in this paper.

One may also see some overestimation of the fusion cross section at above barrier energies for the $^{4}$He+$^{64}$Zn reaction for which the break-up is not expected at all. It is well known that the behavior of the fusion cross section in this energy range is completely determined by the potential parameters. They are chosen here identical for the both reactions in order to avoid additional factors influencing the calculated cross sections. This leads to the close values of the cross sections at hight energies, where they can be approximated by the geometrical value $\pi R_B^2$. This overestimation is not important for our consideration since it is rather small and the aim of this paper is the analysis of the sub-barrier fusion. Thus we did not play with the potential parameters.

As already mentioned there are many evidences for additional sub-barrier fusion enhancement owing to neutron rearrangement with positive $Q$-value both for stable and weakly bound nuclei. However several projectile--target combinations were reported recently ($^{58,64}$Ni+$^{130}$Te \cite{PRL.107.202701}, $^{16,18}$O+$^{76,74}$Ge \cite{PRC.86.044621}, $^{60,64}$Ni+$^{100}$Mo \cite{Fusion_Ni_Mo}) for which no noticeable enhancement of the sub-barrier fusion cross sections was observed in spite of positive $Q$-values for neutron transfer. Here we performed analysis of these fusion reactions within the model formulated above. We found that the model describes well all the experimental data and it also does not predict any significant fusion enhancement for these specific combinations having positive $Q$-values for neutron rearrangement. This sets us studying deeper the mechanism of the intermediate neutron rearrangement in sub-barrier fusion reactions.
    {\setlength{\extrarowheight}{4pt}
    \begin{table}[ht]
        \begin{center}
            \caption{$Q_{xn}$ values (in MeV) of neutron transfers and vibrational properties of the targets. The energies of the first vibrational state are given in MeV.}
            \label{tab_QGG_1}
            \begin{tabular}{p{2cm}|cccc|cc}
              \hline\hline
               Reaction & 1n & 2n & 3n & 4n & $E_{2^{+}}$ &  $\left<\beta_{2}^{0}\right>$ \\
              \hline
              ${}^{40}$Ca+${}^{96}$Zr & +0.51 & +5.53 & +5.24 & +9.64 & 1.75 & 0.080\\
              ${}^{58}$Ni+${}^{130}$Te & +0.58 & +5.89 & +4.92 & +9.23 & 0.84 & 0.118\\
              ${}^{60}$Ni+${}^{100}$Mo & --0.47 & +4.20 & +2.39 & +5.23 & 0.54 & 0.231\\
              ${}^{18}$O+${}^{74}$Ge & --1.54 & +3.74 & --5.85 & --10.40 & 0.60 & 0.283\\
              \hline \hline
            \end{tabular}
        \end{center}
    \end{table}

\begin{figure}
\includegraphics[width=7cm]{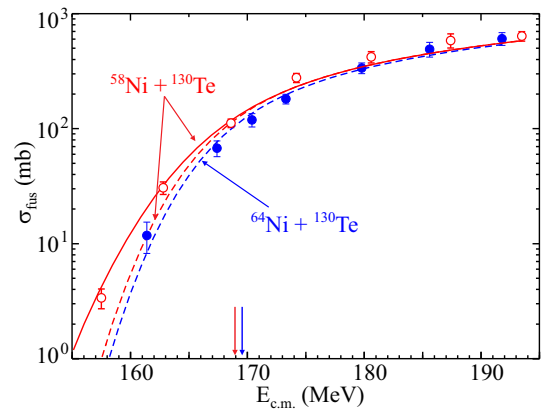}
\caption{(Color online) Fusion cross sections for $^{58}$Ni+$^{130}$Te and ${}^{64}$Ni+${}^{130}$Te. The experimental data  \cite{PRL.107.202701} are shown by open and filled circles, respectively. The solid and dashed curves  show the ECC calculations with and without neutron rearrangement, respectively.}
\label{TeNi}
\end{figure}
\begin{figure}
\includegraphics[width=7cm]{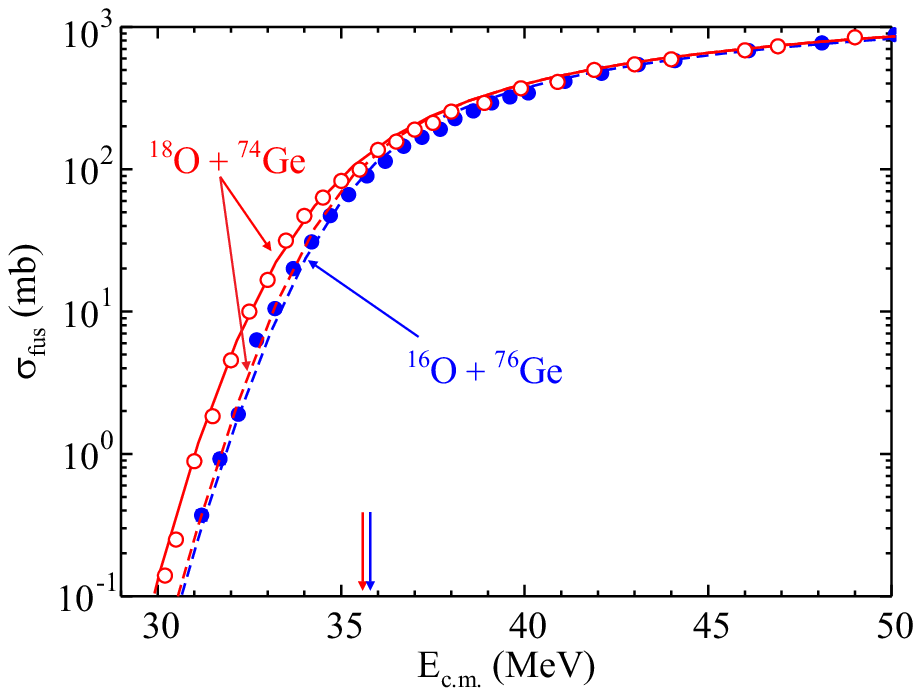}
\caption{(Color online) Fusion excitation functions for $^{18}$O + ${}^{74}$Ge and $^{16}$O + ${}^{76}$Ge. The experimental data  \cite{PRC.86.044621} are shown by open and filled circles, respectively. The solid and dashed curves show the ECC calculations with and without neutron rearrangement, respectively.}
\label{O_Ge}
\end{figure}
\begin{figure}
\includegraphics[width=7cm]{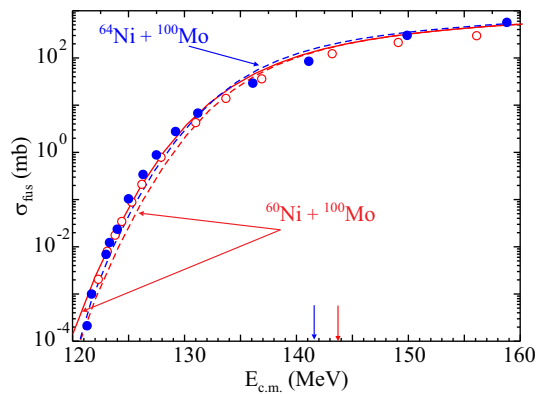}
\caption{(Color online) Fusion excitation functions for ${}^{60}$Ni + ${}^{100}$Mo and ${}^{64}$Ni + ${}^{100}$Mo. The experimental data  \cite{Fusion_Ni_Mo} are shown by open and filled circles, respectively. The solid and dashed curves for show the ECC calculations with and without neutron rearrangement, respectively.}
\label{Ni+Mo}
\end{figure}

The corresponding fusion cross sections and the results of our analysis are shown in Figs.\ \ref{TeNi}, \ref{O_Ge} and \ref{Ni+Mo}. Theoretical calculations agree well with the experimental data. In all the cases neutron rearrangement with positive $Q$-value is taken into account. However only in the case of the $^{58}$Ni+$^{130}$Te reaction with positive $Q_{1n} = +0.58$ MeV and $Q_{2n} = +5.89$ MeV (see Table.\ \ref{tab_QGG_1}) some excess in the sub-barrier fusion cross section is visible as compared with more neutron rich system $^{64}$Ni+$^{130}$Te having all negative $Q$-values for neutron transfers beside $Q_{2n} = +0.55$ MeV (see Fig.\ \ref{TeNi}).

The data do not show any significant effect of neutron rearrangement for the $^{60}$Ni+$^{100}$Mo fusion reaction (having positive $Q$-value neutron transfer from target to projectile, see Table. \ref{tab_QGG_1}) as compered with the $^{64}$Ni+$^{100}$Mo (the only positive value of $Q_{2n} = +0.83$ MeV).
The same takes place for the $^{18}$O+$^{74}$Ge fusion reaction (having positive $Q_{2n} = +3.75$ MeV) as compared with $^{16}$O+$^{76}$Ge (all $Q_{xn}<0$) shown in Fig.\ \ref{O_Ge}.

Note one more that the model used takes into account neutron rearrangement, reproduce quite well the experimental data and also does not predict any sub-barrier fusion enhancement for these specific reactions with positive $Q$-values of neutron transfers. Thus, we have to understand what features of these reactions (properties of colliding nuclei) suppress a gain coming from positive $Q$-value neutron rearrangement (clearly visible in many other cases).

\section{Interplay of nuclear properties and sub-barrier fusion enhancement}\label{Sect4}

In this section different factors influencing the enhancement of the sub-barrier fusion due to neutron rearrangement are discussed. We use here the notation ``enhancement factor'' to characterize the effect of coupling with neutron-transfer channels. A standard way to measure the enhancement factor consists in studying two close projectile-target combinations (to ensure similar fusion barriers and properties of the collective excitations): one with negative $Q$-values (no neutron rearrangement effect) and another one with positive $Q$-values. The difference of the cross sections at sub-barrier region (if it would be observed) must be attributed to the additional coupling with neutron transfer channels. This method is not straightforward because a difference of the fusion cross sections may still appear owing to difference of fusion barriers or the collective excitation properties. However in many cases this method gives a good approximation to the ``real'' enhancement factor, which we define as the ratio of the fusion cross sections obtained with and without account for the coupling to the neutron rearrangement channels. It is clear that such a quantity can be obtained in theoretical calculations only.

\subsection{$Q$-values}

The $Q$-values of neutron transfer as the factor determining the enhancement of the sub-barrier fusion cross section was discussed many times (see, e.g., \cite{PhysRevC.65.011601,ZagrebaevPRC.67.061601}). We repeat here the main points. The statement is that if the system of two colliding nuclei has positive $Q$-values of neutron transfer then one may expect that the sub-barrier fusion cross section will demonstrate enhancement due to neutron rearrangement additional to the one caused by the coupling of the relative motion to vibrational and/or rotational degrees of freedom.
This effect is demonstrated in Fig. \ref{40Ca96Zr_Qxn} for the ${}^{40}$Ca + ${}^{96}$Zr system. The $Q_{xn}$ values for this system are listed in Table \ref{tab_QGG_1}. The experimental fusion cross sections \cite{Timmers1998421} are well-reproduced within the model. However, if one assumes that the $Q_{xn}$ values are twice smaller (dashed curve) then the neutron rearrangement enhancement reduces drastically, and the calculated fusion cross section becomes much closer to the ECC calculations without neutron rearrangement (dotted curve). Opposite, for twice larger $Q_{xn}$ values (dash-dotted curve) the effect is much stronger.

\begin{figure}
\includegraphics[width=7cm]{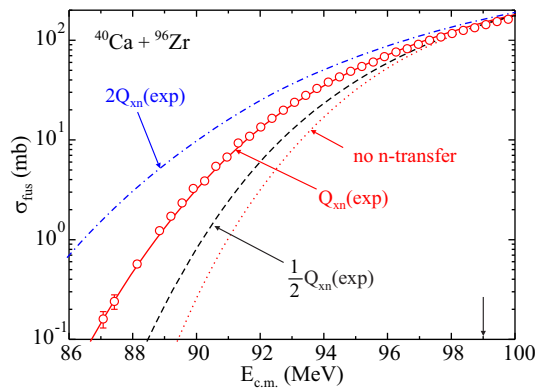}
\caption{(Color online) Fusion excitation functions for ${}^{40}$Ca + ${}^{96}$Zr. The curves show the ECC calculations with and without (dotted curve) account for neutron rearrangement channels. The solid curve is obtained with real $Q_{xn}$ values, the dashed and dash-dotted curves are the model calculations assuming twice smaller and twice larger $Q_{xn}$ values, respectively.}
\label{40Ca96Zr_Qxn}
\end{figure}

The last but not least point is that only the rearrangement of outermost neutrons (normally 1n and 2n transfer channels) may enhance the sub-barrier fusion significantly. This happens because the coupling with neutron rearrangement channels influences the fusion probability only if such rearrangement takes place {\em before} overcoming the Coulomb barrier. Therefore, only a few valent neutrons having largest radii of their wave functions should be taken into account in the analysis of the sub-barrier fusion. Figure \ref{40Ca96Zr_Ntransfer} shows the fusion cross section for the $\rm ^{40}Ca+^{96}Zr$ system calculated within the ECC model with account for different number of neutron rearrangement channels. The $Q_{xn}$ values of the neutron transfer are given in Table \ref{tab_QGG_1} up to 4$n$, the rest values are: $Q_{5n}=+8.42$ MeV and $Q_{6n}=+11.62$ MeV. One may see that the main effect comes from 1n+2n channels. Much smaller but still visible effect is due to rearrangement of 3rd and 4th neutrons, whereas transfer of more neutrons does not influence the sub-barrier fusion probability.
\begin{figure}
\includegraphics[width=7cm]{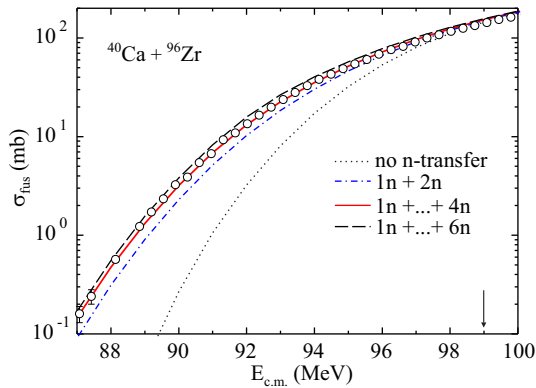}
\caption{(Color online) Fusion excitation functions for ${}^{40}$Ca + ${}^{96}$Zr. The curves show the ECC calculations with account for different number of neutron rearrangement channels (from 0$n$ to 6$n$).}
\label{40Ca96Zr_Ntransfer}
\end{figure}

\subsection{Properties of collective excitations}

The $\rm ^{40}Ca+^{96}Zr$ combination is a typical example of the reaction revealing the strong fusion enhancement at sub-barrier energies due to neutron rearrangement. On the contrary, there are above mentioned projectile-target combinations already studied experimentally that seem to be similar (having the similar $Q$-values of neutron transfers) to the $\rm ^{40}Ca+^{96}Zr$ case but without any significant effect from the neutron rearrangement. They are: $\rm ^{60}Ni+^{100}Mo$ \cite{Fusion_Ni_Mo} and $\rm ^{58}Ni+^{130}Te$ \cite{PRL.107.202701}.

The reason why one observes very different influence of the neutron rearrangement on the sub-barrier fusion for these ``similar'' systems consists in their different vibrational properties. Our statement is that the fusion enhancement due to the neutron rearrangement is larger for the systems having smaller fusion enhancement due to the coupling to collective states.

In the case of fusion reactions of light nuclei the coupling to the collective states of heavy target has rather low impact on the barrier penetrability. This means that the sub-barrier fusion enhancement due to neutron rearrangement should be more pronounced for these reactions in the case of positive $Q$-value neutron transfers.

For the collisions of medium and heavy statically deformed nuclei the coupling to the rotational states always plays a significant role. Thus, for such combinations the effect of neutron rearrangement is expected to be small. For spherical nuclei the sub-barrier fusion enhancement owing to the coupling to the vibrational degrees of freedom can be very different in magnitude depending of the vibrational properties of the reaction partners. Below we will focus just on the analysis of the fusion of spherical nuclei.

\begin{figure}
\includegraphics[width=7cm]{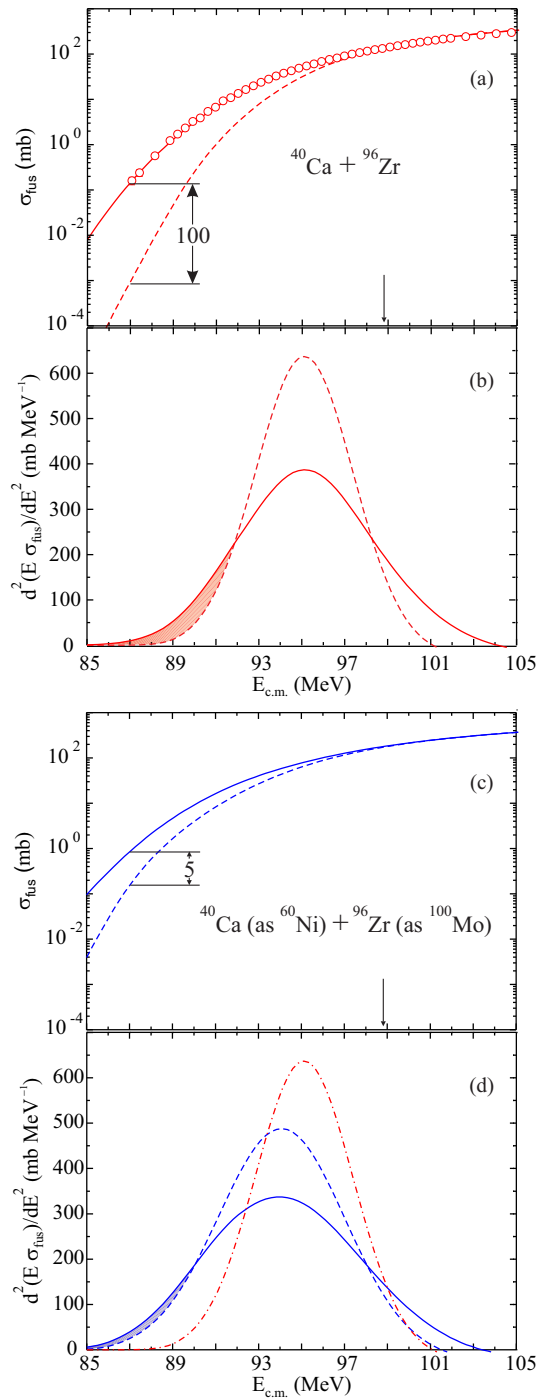}
\caption{The fusion cross sections (a) and (c) and the experimental barrier distribution functions (b) and (d). The solid and dashed curves show the results of the ECC calculations with and without accounting for the neutron rearrangement channels, respectively. The panels (a) and (b) correspond to the $^{40}$Ca+$^{96}$Zr system (the symbols are the experimental data), while the panels (c) and (d) are made for the same system but the vibrational properties of $^{40}$ Ca and $^{96}$Zr are substituted by those of $^{60}$Ni and $^{100}$Mo, respectively. The dash-dotted curve on the panel (d) is the same as the dashed one on the panel (b).}
\label{CaZr_NiMo}
\end{figure}

The influence of vibrational properties of nuclei (their softness) on barrier penetrability can be estimated quantitatively. In the ECC model the coupling to the deformation degrees of freedom is determined by the shape of the empirical dynamic barrier distribution function $f(B)$, see Eqs. (\ref{Total_fusion_CS}) -- (\ref{FB}). In Eq. (\ref{FB}) the effective fusion barrier $B_0$ and the width of the distribution $\Delta B$ are determined by the energy of the first vibrational state $E_{\lambda}$ and the {\em rms} value of zero vibrations $\left<\beta^0_{\lambda}\right>$. The smaller is $\left<\beta^0_{\lambda}\right>$ and the larger is $E_{\lambda}$ (``rigid'' nuclei) the closer is $B_0$ to the barrier for spherical nuclei and the smaller is $\Delta B$, i.e. the narrower is the barrier distribution. The studied enhancement due to the neutron rearrangement is determined by the ratio of the transmission probabilities obtained with and without account for the neutron channels. For simplicity but without loosing generality one may assume that only $1n$ rearrangement channel plays a role. Additionally, we consider only the first term ($l=0$) in the partial-wave decomposition of the fusion cross section. In this case the ``enhancement factor'' can be approximated by
\begin{equation}
\label{Factor}
F=\frac{\sum\limits_{l} (2l+1) \tilde{T}_l}{\sum\limits_{l}  (2l+1) T_l} \sim \frac{\tilde{T}_0}{T_0} \simeq 1 + const\frac{Q_{1n}}{\Delta B}\frac{\exp{\left(-\left[\frac{E-B_0}{\Delta B}\right]^2\right)}}{1+ {\rm erf}\left(\frac{E-B_0}{\Delta B}\right)}.
\end{equation}
This expression is obtained for $Q_{1n}>0$ and under the assumption that $C Q_{1n}^2 \ll 1$, which is quite reasonable for $Q_{1n} \le 5$~MeV. The constant value in (\ref{Factor}) consists of all normalization factors as well as nuclear properties playing the second order role (e.g. the neutron binding energies discussed below).
This approximation of the enhancement factor allows us to conclude that:
\begin{enumerate}
    \item The enhancement factor increases with increase of the $Q_{xn}$ value.
    \item The enhancement is larger for smaller $\Delta B$ values, i.e. for more rigid nuclei.
    \item The enhancement factor increases when the energy goes deeper to the sub-barrier region. For above-barrier energies $F$ tends to unity.
\end{enumerate}

The influence of the collective properties of nuclei on the enhancement of the fusion cross section due to the neutron rearrangement is illustrated in Fig. \ref{CaZr_NiMo}. Figures \ref{CaZr_NiMo} (a) and (b) show the fusion cross sections and the barrier distributions for the $^{40}$Ca+$^{96}$Zr system. The fusion enhancement due to the neutron rearrangement constitutes two orders of magnitude at the energies 12 MeV below the barrier [compare the solid and dashed curves in Fig. \ref{CaZr_NiMo} (a)]. This enhancement is also well seen in the barrier distributions, when one compares the distributions calculated with and without neutron rearrangement at the low-energy region (shadowed area). Both reaction partners are magic spherical nuclei, and, therefore, they are hardly deformed nuclei, since, their first excited states (see Table \ref{tab_QGG_1}) are rather high. According to our conclusion this motivates so large influence of the neutron rearrangement on the sub-barrier fusion.

If now we assume that these nuclei are softer with respect to their deformation than they actually are, and replace the vibrational properties of $^{40}$Ca and $^{96}$Zr by those of $^{60}$Ni and $^{100}$Mo (softer nuclei) the influence of the collective excitations increases [see Fig. \ref{CaZr_NiMo} (c)]. As the result, account for neutron transfer channels gives now additional enhancement factor 5 instead of 100. The effect is well seen in the behavior of the barrier distributions shown in Fig. \ref{CaZr_NiMo} (d). The barrier distribution after change of the vibrational properties of colliding nuclei (the dashed curve) shifts to the lower energies and  becomes wider (smaller value of $B_0$ and larger value of $\Delta B$). As the result the low-energy tails of the distributions with and without account of neutron rearrangement getting very close to each other. The fusion enhancement due to the coupling to the neutron channels is, therefore, much smaller than for the original $^{40}$Ca+$^{96}$Zr system.
Note, that in these calculations we alter only vibrational properties of nuclei and do not change the rest model parameters (potential, charges, masses, binding energies, $Q$-values, etc.).

The performed analysis clearly explains why the effect of the neutron rearrangement is negligible in the above mentioned systems (Fig. \ref{TeNi}, \ref{O_Ge}, \ref{Ni+Mo}), while it is well-pronounced  for the ``similar'' $^{40}$Ca+$^{96}$Zr combination. It is just because of the different properties of their collective excitations shown in Tab. \ref{tab_QGG_1}. The correlation between these properties and the observed fusion enhancement is clearly seen. The smaller is the {\em rms} deformation parameter (and the higher is the energy of the vibrational state) the smaller effect is expected from the coupling with collective states, and, hence, the larger is an influence of the neutron rearrangement on sub-barrier fusion. Weak neutron-channels-caused enhancement for $^{18}$O+$^{74}$Ge is additionally motivated by the fact that the only positive and rather moderate $Q_{xn}$ value corresponds to the 2n channel, while the $Q$-value for the 1n channel is negative. Note, that in Tab. \ref{tab_QGG_1} only the target collective properties are shown, because for the studied systems the coupling to the collective states of the targets is more important and has the largest effect. However in calculations both the collective properties of targets and projectiles are included.


\subsection{Neutron binding energies}

\begin{figure}[t]
\includegraphics[width=8cm]{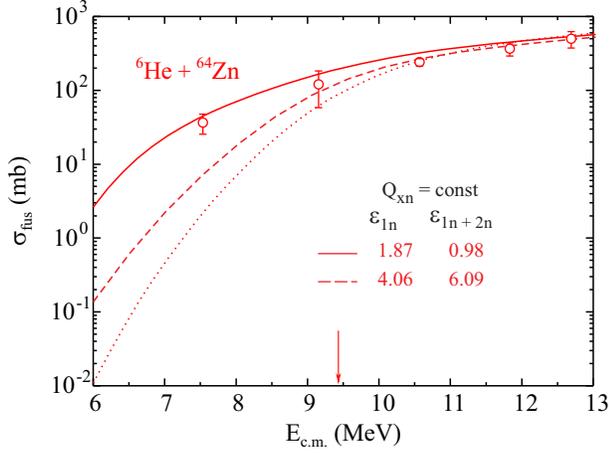}
\caption{(Color online) Fusion cross sections for ${}^{6}$He + ${}^{64}$Zn reaction. The data (symbols) are taken from Ref. \cite{JPhys.282.012014}. The curves are the ECC calculations with (solid and dashed) and without (dotted) neutron rearrangement. The solid curve corresponds to real neutron binding energies in $^{6}$He. The dashed curve shows the model calculations assuming larger neutron binding energies in $^{6}$He (the same as in $^{9}$Li). The binding energies are shown in MeV.}
\label{6He64Zn}
\end{figure}
\begin{figure}[t]
\includegraphics[width=8cm]{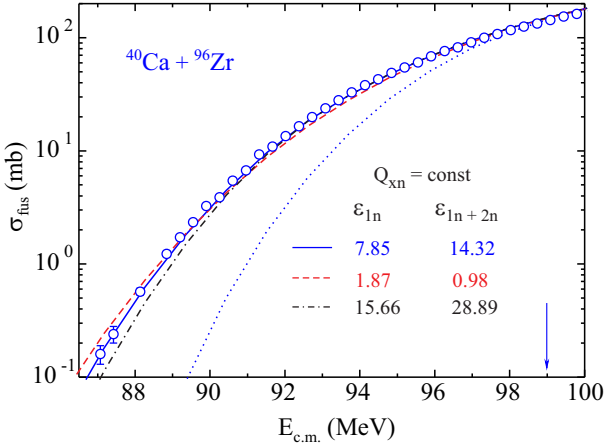}
\caption{(Color online) Fusion cross sections for ${}^{40}$Ca + ${}^{96}$Zr reaction. The data (symbols) is taken from Ref.  \cite{Timmers1998421}. The curves are the ECC calculations with (solid, dashed, and dash-dotted) and without (dotted) neutron rearrangement. The solid curve corresponds to real neutron binding energies in $^{96}$Zr. The dashed and dash-dotted curves show the model calculations assuming smaller (as in $^{6}$He) and larger (as in $^{16}$O) neutron binding energies in $^{96}$Zr, respectively. The binding energies are shown in MeV.}
\label{CA96Zr}
\end{figure}
\begin{figure}[t]
\includegraphics[width=8cm]{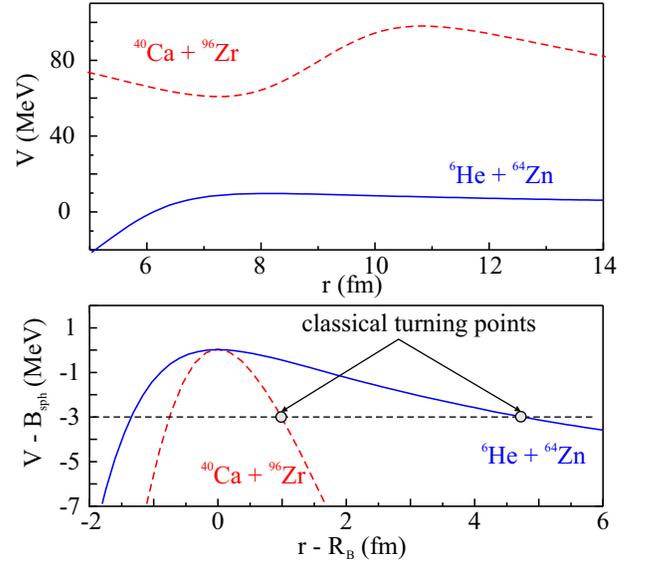}
\caption{The nucleus-nucleus potential energies for the ${}^{6}$He+${}^{64}$Zn and ${}^{40}$Ca+${}^{96}$Zr systems
and classical turning points at $E_{\rm c.m.} = B-3$~MeV.}
\label{pot}
\end{figure}

In order to clarify the role of the neutron binding energies we performed the following calculations. The binding energies of two valent neutrons were varied simultaneously in the target and projectile preserving all the other properties ($Q$-values, potentials, etc.) unchanged. In what follows we will discuss the neutron binding energies in the ``donor'' nucleus only. The influence of the neutron binding energies on the sub-barrier fusion of light weakly bound nuclei is shown in Fig. \ref{6He64Zn}. One may see that at energies $\sim 3$ MeV below the barrier the total effect of the neutron rearrangement constitutes about two orders of magnitude (the solid curve as compared to the dotted one). If in the calculations one uses more bound neutrons in $^{6}$He (the same values of $\varepsilon_{xn}$ as in ${}^{9}$Li) but all other properties (potential, $Q$-values, etc.) keeps unchanged, then the effect reduces to one order of magnitude.

However in sub-barrier fusion of heavy nuclei the binding energy of transferred neutron has almost no impact on fusion enhancement!
One may see in Fig. \ref{CA96Zr} that in the case of ${}^{40}$Ca + ${}^{96}$Zr sub-barrier fusion cross sections shift only a little if one assumes neutrons in ${}^{96}$Zr less bound (in ${}^{6}$He) or twice stronger bound (as in ${}^{16}$O).

The reason why the neutron binding energy plays more important role in fusion of light nuclei becomes clear from Fig. \ref{pot}, where two nucleus-nucleus potentials are shown for the ${}^{6}$He + ${}^{64}$Zn and ${}^{40}$Ca + ${}^{96}$Zr systems. For the lighter projectile  the Coulomb barrier is lower but wider and the classical turning point corresponds to larger distance between nuclear surfaces than for the heavier one due to the smaller $Z_1Z_2$ Coulomb factor. This discrepancy increases with decreasing energy below the Coulomb barrier. The neutron binding energy determines the ``compactness'' of its wave function and hence the neutron transfer probability (\ref{ak}), which is also dependent on the position of the turning point [``$D-D_0$'' factor in (\ref{ak})]. Thus, the neutron transfer probability decreases much faster with increasing the binding energy in the case of lighter nuclei.

On the other hand, for the lighter systems the same gain in energy (determined by the $Q$-value) has larger relative influence on each item in the sum for the transmission coefficient (\ref{Tl_HW_MOD}) than for heavier one because of the lower Coulomb factor. This explains why the largest enhancement of the sub-barrier fusion cross sections owing to neutron rearrangement is expected (and observed experimentally) for the fusion of light weakly bound nuclei having large positive $Q$-values for neutron transfer.


\section{Conclusions}

The role of neutron rearrangement channels in near-barrier fusion reactions is studied within the ECC model. It is shown that the model reproduces available experimental data quite well. It also gives close results to the QCC model when the neutron rearrangement does not play a role. In contrast with the generally accepted opinion, we found that the positive $Q$-values for the neutron transfer is not the only factor determining enhancement of sub-barrier fusion probability. A noticeable additional enhancement of sub-barrier fusion cross section (beside those caused by coupling to collective degrees of freedom) can be expected in the following cases.
\begin{enumerate}
    \item When the system has large positive $Q$-values for transfer of one or/and two neutrons. The role of $xn$-channels is negligible for $x>4$.
    \item When the coupling to the collective states is not so much important at sub-barrier energies. This is always the case when one of the reaction partners is a light nucleus. In all other cases this is realized for the fusion of spherical nuclei having high energies of the first vibrational state and small values of the {\em rms} dynamical deformation. The largest effect from the neutron rearrangement one should expect for spherical nuclei having magic or nearly magic numbers of protons or/and neutrons.
\end{enumerate}
Additional enhancement of the sub-barrier fusion cross sections takes place for light neutron-rich weakly bound nuclei due to smaller binding energies of valent neutrons.

\begin{acknowledgments}
The work was supported by the Russian Foundation for Basic Research under Grant No. 12-02-01325 and No. 13-07-00714-a.
\end{acknowledgments}

\end{document}